\documentclass[prl,twocolumn,superscriptaddress]{revtex4-1}
\usepackage{graphicx}
\usepackage{amsmath}
\usepackage{dcolumn}
\usepackage{color}
\usepackage{epstopdf}
\usepackage[caption=false]{subfig}
 
\newcommand\lb{\langle}
\newcommand\rb{\rangle}

\newcommand\emfb{\overline{\mbox{\boldmath ${\cal E}$}} {}}
\newcommand\emf{\overline{\mbox{${\cal E}$}} {}}

\begin{document} 

\title{Minimalist large scale dynamo from shear-driven inhomogeneity}

\author{F. Ebrahimi}
\affiliation{Princeton Plasma Physics Laboratory, and Department of Astrophysical Sciences, Princeton University NJ, 08544}
\author{E.G. Blackman}
\affiliation{Department of Physics and Astronomy, University of Rochester, Rochester, NY 14627, USA} 

\date{\today}
  
\begin{abstract}
We show that non-axisymmetric, non-helical perturbations in an unstratified shear flow 
produce a shear-plane averaged electromotive force (EMF)  proportional to a spatially dependent kinetic helicity. This new "shear-driven $\alpha$-effect"  can   amplify  shear-plane averaged magnetic  fields, but is undiscoverable when invoking the   homogeneity assumption for  dynamo coefficients.
It  is a candidate mechanism underlying  large scale field generation in astrophysical, laboratory, and simulated shear flows, and highlights the essentiality of  correctly averaging  dynamo coefficients and  fields.

\end{abstract}
\maketitle

\textit{Introduction--}  A plethora of astrophysical objects--including planets, stars, accretion disks, and  galaxies--possess large scale magnetic fields. 
Understanding the  in situ  sustenance of these large scale   fields has been a long-standing subject of research  tackled with mean field dynamo models that use suitably chosen spatial, temporal, or ensemble averages.  Global dynamo models  aimed  toward  astrophysical realism can be distinguished from  "proof-of-principle"  theory aimed at  identifying key underlying physical principles.   The latter  have long helped  to push our understanding forward.
  
 The question arises as to
 whether even  minimalist shear flows can amplify large scale fields for a suitable chosen averaging procedure.   For systems subjected to different sources of statistically non-helical fluctuations or turbulence with shear,   large scale dynamos of the (i) traditional $\alpha-\Omega$ type \cite{moffatt,Parker1979}   have been considered \cite{Brandenburg1995,Guan2011},  as well as dynamos that involve  (ii) shear plus fluctuations
\cite{vishniac97,Blackman1998,Yousef2008,Heinemann2011}, or  (iii) the shear current effect from negative diffusion \cite{shearcurrent, Kleeorin2003+,sridhar2014,Squire2015+}.
Each of these dynamos were first conceptually introduced in their  original form
using some minimalist version, typically using  the first  order smoothing  or quasi-linear approximation before  generalizations that included turbulent closures.

 In this same spirit,  we uncover a new shear-driven large scale dynamo mechanism.   We show analytically that  the shear-plane averaged electromotive force (EMF)  for an unstratified non-rotating shear flow  subjected to 2-D perturbations 
leads to  a  spatially dependent dynamo coefficient that produces large scale dynamo growth.
The large-scale dynamo here is defined as exponential  growth  of a shear-plane averaged field. 
The  shear-flow breaks the symmetry and causes a correlation of magnetic and flow fluctuations.  

For  clarity, we first present the calculation of the EMF rigorously, before   discussing the  conceptual connections to previous mean field dynamo theories and how a common flaw in  averaging procedure hides this mechanism.

\textit{Shear flow and perturbations--} We consider an unstratified shear flow in  non-rotating local Cartesian coordinates $(x,y,z)$ with a linear shear velocity of $\bf \bar V_0$ = $\overline V_y(x) \bf{\hat y}  $, and a
weak initial seed field $\bf \bar B_0$ = $B_{0z}{\bf \hat z} + B_{0y}{\bf \hat y}$. We assume  perturbed 
 velocity and magnetic fields
  of the form  $\pmb{\xi}(x, y, z, t) = [\xi_x(x),\xi_{y}(x),\xi_z(x)]\exp{(\gamma_c t - ik_yy+ik_z z)}$, where, $\gamma_c = \gamma +i\omega_r$, $\gamma$ is the real time growing or decaying part and $\omega _r$ is the oscillatory part of the perturbations.  All variables are decomposed as $\xi(x,y,z,t)=\lb \xi(x,t)\rb+\widetilde \xi(x,y,z,t)$, where $\lb \xi \rb$ (or $\bar{\xi}$) is the mean  component, and $\widetilde \xi $ is the fluctuating component.   We call $y$ and  $x$ directions  ``azimuthal" and ``radial" respectively as if the Cartesian system were locally embedded in a rotating shear flow, even though we ignore rotation.\footnote{In the notation of conventional planar shear flows, our directions are: $y=$ streamwise, $z=$ spanwise and $x=$ wall-normal.} 
  Mean quantities  indicated  by brackets $\lb \rb$ or overbars are  shear-plane (azimuthally and axially) averaged  but remain dependent on $x$.
  The spatial variation of fluctuations (specified by $k_y$ and $k_z$) here are arbitrary, and could be smaller than or close to the scale of the system.

  In this non-rotating  unstratified system, the momentum equation $\rho \frac {\partial \textbf V }{ \partial t } =
- \rho \textbf V . \nabla\textbf V +
 \textbf J \times
 \textbf B - \nabla P$
 and the induction equation $\partial_t {\bf B}=\nabla \times ({\bf V}\times {\bf B})$
are linearized in the incompressible limit to give:
\begin{equation}
\begin{split} 
\overline{\gamma}\widetilde{V}_x  &=i  (\mathbf{k} \cdot {\mathbf{\bar{B}_0}}) \widetilde{B}_x -\frac{\partial X}{\partial x}  \\
\overline{\gamma}\widetilde{V}_{y} + \overline V_y^{\prime}(x)\widetilde{V}_x &=  i (\mathbf{k} \cdot \mathbf{\bar{B}_0)} \widetilde{B}_{y} + ik_y X \\
\overline{\gamma}\widetilde{V}_{z} &=i (\mathbf{k} \cdot \mathbf{\bar{B}_0}) \widetilde{B}_{z}  - ik_z X 
\end{split}
\label{eq:momentum}
\end{equation} 
and
\begin{equation}
\begin{split} 
\overline{\gamma} \widetilde{B}_x&= i (\mathbf{k} \cdot {\mathbf{\bar{B}_0}}) \widetilde{V}_x\\
\overline{\gamma} \widetilde{B}_{y}&= \overline V_y^{\prime}(x) \widetilde{B}_x + i (\mathbf{k} \cdot {\mathbf{\bar{B}_0}}) \widetilde{V}_y\\
 \overline{\gamma} \widetilde{B}_{z}&= i (\mathbf{k} \cdot {\mathbf{\bar{B}_0}}) \widetilde{V}_z,
\end{split}
\label{eq:induction}
\end{equation}
where $X = \widetilde{P} + \widetilde{\bf B}\cdot {\bf B}_{0} $;   primes denote derivative in the $x$ direction, ($e.g. \ \partial/\partial x$); and $\vartheta (x) = k_y \overline V_y(x) -\omega_r$, $\overline{\gamma} = \gamma - i \vartheta (x)$.
We ignore the 
$x$ (radial) dependence of  fluctuations ($k_x=0$). Combining Eqs~\ref{eq:momentum} and ~\ref{eq:induction}  gives:
\begin{equation}
  \begin{split}
  \frac{\overline{\gamma}}{M^2}(1+M^2) \widetilde{V}_y&= -(1+\frac{1}{M^2}) \overline V_y^{\prime}(x)\widetilde{V}_x + ik_y X\\
  \frac{\overline{\gamma}}{M^2}(1+M^2) \widetilde{V}_z&= -i k_z X
  \end{split}
  \label{3}
  \end{equation}
where  $M\equiv \frac{\overline{\gamma}^2}{F^2}$, and $F\equiv \textbf k\cdot \bf \bar B_0$. The two equations of (\ref{3}) can be combined with the  incompressibility equation $\widetilde V_z= k_y \widetilde{V}_y/k_z$  and $ k^2=k_z^2+k_y^2$ to give
\begin{equation}
  \widetilde{V}_y = \frac{-k_z^2 \overline V_y^{\prime}(x) \widetilde{V}_x}{k^2 \overline{\gamma}}.
  \label{eq:vy}
  \end{equation}

To uncover the essential physics, we employ a quasilinear approach, in which all the 
 correlation 
terms are constructed from the linear perturbations above. Below, we present all the steps to calculate a shear-driven large scale dynamo directly from the EMF. 

\textit{ EMF calculation and dynamo growth--} The shear-plane averaged quasilinear electromotive force
can now be constructed from the  linear perturbations of magnetic field and velocity. 
We calculate
the  vertical EMF component  
$\emf_z= \lb \widetilde {\bf V} \times \widetilde {\bf B}\rb_{z}={1\over 2} Re(\widetilde{B}_{y} \widetilde V_x^{*} - 
\widetilde V_{y}\widetilde{B}_x^{*})$.
Since  $\widetilde B_x^{*}$ and $\widetilde V_y$ are out of phase [i. e. with $\mathrm{exp}(i\frac{\pi}{2})$ phase difference],  the second term  on the right satisfies $Re[\widetilde V_{y} \widetilde{B}_x^{*}]= Re\left[(\frac{-k_z^2 \overline V_y^{\prime}(x) \widetilde{V}_x}{k^2 \overline{\gamma}})(\frac{-iF\widetilde V_x^{*}}{\overline\gamma^{\ast}})\right] = Re\left[\frac{ik_z^2F \overline V_y^{\prime}(x) |V_x|^2}{k^2(\gamma^2+\vartheta^2)}\right]=0$, using Eqs.  (\ref{eq:induction}) and (\ref{eq:vy}). However  $\widetilde B_y$ and $\widetilde V_x^{*}$ are in phase, so the first term on the right of $\emf_z$  satisfies  $Re[\widetilde{B}_{y} \widetilde V_x^{*}] = Re\left[\frac{iF \overline V_y^{\prime}(x)}{\overline{\gamma}^2}|\widetilde V_x|^2-\frac{iF \overline V_y^{\prime}(x)k_z^2}{k^2 \overline{\gamma}^2}|\widetilde V_x|^2\right]$. Using
$\overline{\gamma}^2 = (\gamma^2 -\vartheta^2 -2i\gamma \vartheta)$, we then arrive at
  \begin{equation}
    \emf_z(x) = \lb \widetilde {\bf V} \times \widetilde {\bf B}\rb_{z}= \frac{k_y^2}{k^2}\left(\frac{-\gamma  \vartheta(x) \overline V_y^{\prime}(x)}{[\gamma^2+\vartheta(x)^2]^2}\right)(\mathbf{k} \cdot {\mathbf{\bar{B}_0}})|\widetilde V_x|^2.
    \label{eq:emf}
    \end{equation}

  This expression reveals three  minimum requirements for a nonzero planar averaged $\emf_z$: (i) a linear shear $\overline V_y(x)^{\prime} \neq 0$;   (ii) non-axisymmetric ($k_y\neq 0$) with $k_z\neq 0$ finite amplitude perturbations  $|\widetilde V_x|$; (iii)  $\gamma$ in the EMF requires a real part of the temporal variation of the perturbations, 
  namely that perturbations must grow or decay (c.f. oscillations are insufficient) to give a finite EMF.
  
  Without resistivity, the shear-plane averaged azimuthal field $\langle B_y\rangle$ can thus be generated according to the  shear-plane averaged  induction equation
\begin{equation}
\frac{\partial \overline{\textbf B}_{y}}{\partial t} = - \frac{\partial \emf _z}{\partial x}
 + (\overline{\textbf B} \cdot \nabla) \overline{\textbf V}|_{y} - (\overline{\textbf V} \cdot \nabla) \overline{\textbf B}|_{y}.
\label{eq:induction1}
\end{equation}
Since there is neither a mean radial  magnetic field
${\overline B}_x$,  nor velocity field ${\overline V}_x$,
 the second term on the right, which is 
the traditional "$\Omega$" effect, along with the third term on the right  both vanish in
Eq. (\ref{eq:induction1}).
A  large scale azimuthal magnetic field, $\overline B_{y} = \langle B_y\rangle \sim- {1\over \gamma_{\mathrm{gr}}} \frac{\partial \emf_z}{\partial x}$, can be directly generated via the spatially dependent $\emf_z$ (Eq.~\ref{eq:emf}).
The growth of the large-scale field $\overline B_{y}$, even from zero ($\overline B_{y} (t=0)=0$), is therefore exponential with  a growth rate of $\gamma_{\mathrm{gr}} = 2 \gamma$, where $\gamma$ is the real part of exponentially varying perturbations [$\emf_z (x,t)= \lb \widetilde {\bf V} \times \widetilde {\bf B}\rb_{z} \mathrm{exp} (2 \gamma t)$]. The temporal variation of $\overline B_{y}$ is directly related to the real time variation of perturbations.  

The vertical field $\overline B_z$  can also grow exponentially  
according to $\frac{\partial \overline{\textbf B_{z}}}{\partial t} =  \frac{\partial \emf_{y}}{\partial x}$ (other terms on the right  side of the induction equation vanish due to  shear-plane averaging).  This growth  could occur starting from nonzero vertical seed field with $\overline B_{0y}=0$ or 
a small nonzero seed azimuthal field $\overline B_{0y}$ with $\overline B_{z} (t=0)=0$. 
Following the steps above, the azimuthal EMF can similarly be obtained as, $\emf_{y } = \frac{k_z}{k_y}  \emf_z$. We can now obtain the total large-scale
(which survived shear-plane
averaging) magnetic energy generated in terms of EMF,
  \begin{equation}
\langle B(x,t)\rangle^2 = \overline B_z^2+ \overline B_y^2 = \frac{1}{4\gamma^2} \frac{k^2}{k_y^2} \left[\frac{\partial \emf_z}{\partial x}\mathrm{exp}(2\gamma t)\right]^2
\label{eq:energy}
  \end{equation}
where $\emf_z$ is given by Eq.~\ref{eq:emf}. This \textit{large-scale field energy} is  generated from non-zero fluctuations in a shear-flow. 

\textit{Helicity densities--} To facilitate   comparison of this  large scale dynamo with  conventional mean field dynamos,
we examine the  shear-plane averaged kinetic and current helicity densities  associated with the
fluctuations.

The shear-plane $(y,z)$
 averaged kinetic helicity is $\lb\widetilde \textbf V \cdot \nabla \times \widetilde \textbf V\rb= \frac{1}{4}[(\widetilde{\omega}_x+\widetilde{\omega}_x^{\ast})(\widetilde{V}_x+\widetilde{V}_x^{\ast})+(\widetilde{\omega}_y+\widetilde{\omega}_y^{\ast})(\widetilde{V}_y+\widetilde{V}_y^{\ast})+ (\widetilde{\omega}_z+\widetilde{\omega}_z^{\ast})(\widetilde{V}_z+\widetilde{V}_z^{\ast})]$= $\frac{1}{2}Re[-ik_y(\widetilde{V}_z\widetilde{V}_x^{\ast}-\widetilde{V}_x\widetilde{V}_z^{\ast}) + i k_z(\widetilde{V}_x\widetilde{V}_y^{\ast}-\widetilde{V}_y\widetilde{V}_x^{\ast})=\frac{1}{2}Re\left[\frac{i (k_y^2+k_z^2)}{k_z}(\widetilde{V}_x\widetilde{V}_y^\ast -  \widetilde{V}_y\widetilde{V}_x^\ast)\right]$, where $\widetilde \omega = \nabla \times \widetilde \textbf V$. Using Eq.(\ref{eq:vy}), we then obtain
  \begin{equation}
  \begin{split}
\lb\widetilde \textbf V \cdot \nabla \times \widetilde \textbf V\rb  
=\frac{1}{2}Re\left[k_z \left(\frac{1}{\overline{\gamma}} - \frac{1}{\overline{\gamma}^\ast}\right) i \overline V_y^{\prime}(x) |V_x|^2\right]
    \\
   = \frac{ -k_z \vartheta(x) \overline V_y^{\prime}(x)}{[\gamma^2+\vartheta^2]}|\widetilde{V}_x|^2.
    \end{split}
    \label{eq:kinetic}
\end{equation}
Equation (\ref{eq:kinetic}) shows that the kinetic helicity has a spatial dependence on $x$ from the  variation  of the mean flow.

We next  calculate the current helicity similarly,  which can be written as
$\lb\widetilde{\bf J}\cdot \widetilde{\bf B}\rb =\frac{1}{4}[(\widetilde{J}_x+\widetilde{J}_x^{\ast})(\widetilde{B}_x+\widetilde{B}_x^{\ast})+(\widetilde{J}_y+\widetilde{J}_y^{\ast})(\widetilde{B}_y+\widetilde{B}_y^{\ast})+ (\widetilde{J}_z+\widetilde{J}_z^{\ast})(\widetilde{B}_z+\widetilde{B}_z^{\ast})]$ = $\frac{1}{2}Re[-ik_y(\widetilde{B}_z\widetilde{B}_x^{\ast}-\widetilde{B}_x\widetilde{B}_z^{\ast}) + i k_z(\widetilde{B}_x\widetilde{B}_y^{\ast}-\widetilde{B}_y\widetilde{B}_x^{\ast})$, which reduces to $\lb\widetilde{\bf J}\cdot \widetilde{\bf B}\rb  = \frac{1}{2}Re\left[\frac{i (k_y^2+k_z^2)}{k_z}(\widetilde{B}_x\widetilde{B}_y^\ast -  \widetilde{B}_y\widetilde{B}_x^\ast)\right]$. Using Eq. (\ref{eq:induction}) for $\widetilde B_y$, we  then obtain,
      \begin{equation}
       \begin{split}
       &
   \lb \widetilde{\bf J}\cdot \widetilde{\bf B}\rb  =
  \frac{1}{2} Re\left[\frac{k^2}{k_z} (\frac{1}{\overline{\gamma}^\ast} - \frac{1}{\overline{\gamma}}) (i \overline V_y^{\prime}(x) |B_x|^2)\right]\\&
    + \frac{1}{2} Re\left[ \frac{ik^2F^2}{k_z\overline{\gamma} \overline{\gamma}^{\ast}}[\widetilde{V}_x\widetilde{V_y}^\ast - 
     \widetilde{V}_y\widetilde{V_x}^\ast]\right],
     \end{split}
\end{equation}
where the last term is related to kinetic helicity $ \lb\widetilde \textbf V \cdot \nabla \times \widetilde \textbf V\rb) = \frac{1}{2} Re \left[\frac{i k^2}{k_z}(\widetilde{V}_x\widetilde{V_y}^\ast -  \widetilde{V}_y\widetilde{V_x}^\ast)]\right]$, and the total contribution can be written as 
\begin{equation}
  \begin{split}
    & \lb\widetilde{\bf J}\cdot \widetilde{\bf B}\rb = \frac{1}{2}Re\left[\frac{k^2}{k_z} (\frac{2 \vartheta(x)\overline V_y^{\prime}(x)}{(\gamma^2+\vartheta^2)}) |\widetilde{B}_x|^2\right] \\&+\frac{F^2}{(\gamma^2+\vartheta^2)}\lb\widetilde \textbf V \cdot \nabla \times \widetilde \textbf V\rb.    \end{split}
  \label{eq:total}
\end{equation}

  We now use Eq.~(\ref{eq:kinetic}) in Eq. (\ref{eq:total})  to obtain
\begin{equation}
  \lb\widetilde \textbf J \cdot \widetilde \textbf B\rb = \frac{k_y^2}{k_z} \left(\frac{ \vartheta(x)\overline V_y^{\prime}(x)}{[\gamma^2+\vartheta^2]^2} \right)(\mathbf{k} \cdot {\mathbf{\bar{B}_0}})^2 |\widetilde{V}_x|^2.
  \label{eq:total2}
\end{equation}  
Combining Eqs. (\ref{eq:kinetic}) and (\ref{eq:total2}) gives
\begin{equation}
  \lb\widetilde \textbf J \cdot \widetilde \textbf B\rb  = -\left[\frac{k_y^2 (\mathbf{k} \cdot {\mathbf{\bar{B}_0}})^2}{k_z^2(\gamma^2+\vartheta^2)}\right]\lb\widetilde \textbf V \cdot \nabla \times \widetilde \textbf V\rb.
  \label{eq:helicity2}
  \end{equation} 
 Eqs. (\ref{eq:helicity2}) and  (\ref{eq:kinetic}) show that the EMF 
(Eq.~\ref{eq:emf}) can  be expressed in terms of {\it either} kinetic or current helicities:
\begin{equation}
  \begin{split}
    &\emf_z= \lb \widetilde {\bf V} \times \widetilde {\bf B}\rb_{z}= -\frac{k_z \gamma} {k^2F} \lb\widetilde \textbf J \cdot \widetilde \textbf B\rb \\&= \frac{\gamma} {k^2}\left[\frac{(\mathbf{k} \cdot {\mathbf{\bar{B}_0}})k_y^2}{k_z(\gamma^2+\vartheta^2)}\right]\lb\widetilde \textbf V \cdot \nabla \times \widetilde \textbf V\rb.
  \end{split}
  \label{eq:emf2}
\end{equation}

\textit{Fluctuations with {$k_x\neq 0$}} -- 
 We show here that the large-scale dynamo
above still persists with a radial dependence of fluctuations. 
Following the steps above,  the radial dependence of the fluctuations augments  Eq. (\ref{eq:emf}). For  triply periodic boundaries, 
the variation can be written in terms of $ik_x$, and the EMF due to 
$k_x\neq 0$ becomes $\emf_{z{(k_x\neq 0)}} = \frac{k_x k_y  F \vartheta(x)}{k^2[\gamma^2+\vartheta(x)^2]} |\widetilde V_x|^2$. We find,
 \begin{equation}
  \begin{split}
  \emf_{z\mathrm{(total)}} (x) = \emf_{z{(k_x=0)}} + \emf_{z{(k_x\neq 0)}} &= \\
   \left[\frac{-\gamma k_y \overline V_y^{\prime}(x)}{(\gamma^2+\vartheta(x)^2)}+k_x \right]\frac{k_y (\mathbf{k} \cdot {\mathbf{\bar{B}_0}}) \vartheta(x)}{k^2(\gamma^2+\vartheta(x)^2)} |\widetilde V_x|^2.
   \end{split}
   \label{general}
  \end{equation}

For finite wall boundaries total EMF becomes  $\emf_{z\mathrm{(total)}}= \emf_{z{(k_x=0)}} + \frac{k_y \gamma F}{k^2[\gamma^2+\vartheta(x)^2]} \widetilde V_x^{\prime}\widetilde V_x^{\ast}$, where $\emf_{z{(k_x=0)}}$ is
from  Eq. (\ref{eq:emf}) and the last term arises  
for finite $\widetilde V_x^{\prime}$. 
 For either boundary case and growing fluctuations ($\gamma >0$),  the EMF
 terms (for $k_x=0$ and $k_x\ne 0$) in Eq. (\ref{general})
 have different functional dependencies. These contributions add,  but the shear flow contribution 
 dominates.
 For decaying fluctuations ($\gamma<0$), the contributions can have different signs and the total EMF could be quenched. 
 For $\gamma =0$, and periodic boundaries, the total EMF can still be finite if  $k_x\neq 0$.

 Solutions with $k_x\ne 0$
 are related to the divergence of the helicity flux $\emf_{{(k_x\neq0)}} \cdot {\bf B} \sim \nabla \cdot 
\langle \widetilde {\bf A} \times (\overline {\bf V} \times \widetilde {\bf B})\rangle$.

\textit{Relation to Conventional Mean Field Dynamos--} 
In conventional mean field  theory, 
the  EMF is  expanded as a sum of tensor products of transport coefficients with linear functions of the  mean magnetic field and its derivatives \citep{moffatt,Parker1979}. The $\alpha_{ij}$ tensor conventionally indicates the coefficient proportional to the term linear the $0th$ derivative of the mean field. For kinematic theories $\alpha_{ij}=\alpha_{ij}({\widetilde {\bf V}})$
but  in general $\alpha_{ij}=\alpha_{ij}({\widetilde {\bf V}},{\widetilde {\bf B}})$.
If the initial field is uniform and equal to $B_z$, then
$\emf_z = \alpha(x) \overline{B}_z$, where $\alpha(x)=\alpha_{zz}(x)$.\\

We can connect Eqs. (\ref{eq:emf}) and  (\ref{eq:emf2}) to this formalism by inspection, and extract the
$\alpha$-coefficient as
\begin{equation}
  \begin{split}
  \alpha(x) &= \left(\frac{-\gamma  \vartheta(x) \overline V_y^{\prime}(x)}{[\gamma^2+ \vartheta(x)^2]^2}\right)\frac{k_z k_y^2}{k^2} |\widetilde V_x|^2\\
 &= \left(\frac{\gamma k_y^2}{k^2(\gamma^2+\vartheta^2)}\right)\lb\widetilde \textbf V \cdot \nabla \times \widetilde \textbf V\rb.
  \end{split}
  \label{eq:alpha}
\end{equation}
The   spatial non-uniformity of this shear-plane-averaged $\alpha$ coefficient is directly related to the flow shear.
 We see that  standard approaches which presume homogeneity of the $\alpha$ coefficient
 {\it a priori}  \citep{moffatt,Parker1979} would be unable to discover the shear-plane averaged dynamo that we have identified since
  it depends on the inhomogeneity of $\alpha$.
  
Our calculations also differ from those  exhibiting large scale dynamos
from non-helically forced turbulent systems with superimposed linear flow shear
\cite{Yousef2008}.
For these systems, a fluctuating $\alpha$-effect~\cite{vishniac97} using mean field theory may be  effective in explaining the results.~\cite{Heinemann2011,sridhar2014} However, the planar averaging  in these calculations is taken over  $x,y$
directions, and the mean field sought is a function of $z$. 
That averaging procedure eliminates any possibility for the dynamo coefficients to depend on the the spatial direction in which the mean velocity varies, in turn eliminating any possibility to discover the systematic $\alpha$-effect 
we  have identified in our calculation above.

A further difference from  standard large scale dynamos
in turbulent systems is  that our calculation applies only where  fluctuation magnitudes are no larger than that of   the initial seed mean field, $\tilde V \sim \tilde B \le B_0$.  Were it not for the presence of shear, these fluctuations would be purely wave-like and the EMF would vanish.  The free energy for the large scale field to grow--even when it is already strong compared to the fluctuations--is supplied by the shear.
We have assumed  that the background shear flow is  not influenced
by the growing field, and thus have not considered the back-reaction on the shear flow.

Contexts where the conditions for our minimalist dynamo  
arise  include:
(i) coronal loops, where the field of the loop
provides the initial $B_0$  subjected to perturbations and  shear perpendicular to this field at the base. The twisting of the foot-points could produce instabilities that provide
 EMF-sustaining fluctuations, which in turn sustain
a large scale helical field along the loop;  
(ii) laboratory plasmas  nonaxisymmetric magnetic fluctuations ~\cite{ebrahimi2016dynamo} (arising from reconnecting or Kelvin-Helmholtz instabilities), combined with local shear flows (strong azimuthal flows with radial dependency) could cause a phase shift that  generates an EMF and facilitates global relaxation; (iii)
generation of surface zonal fields in planetary systems from an initially radial perturbed field
penetrating a strong zonal shear flow;  (iv) when rotation is included,
the generalized version reveals that large scale field growth in  shearing box MRI simulations  precedes the onset of mode coupling and small scale turbulent fields. 
\cite{,EB_2016,bhat2016}; (iv) also for rotating systems, if an accretion disk is threaded by a strong vertical field, our mechanism can create toroidal from poloidal fields and thereby help to initiate a Poynting flux mediated jet.  

The dynamo in our quasi-linear calculation 
 does not depend on  dissipation. 
 This  contrasts  Ref. \cite{Keinigs1983}, which shows that without shear, dissipation is required to establish a phase correlation to sustain an EMF for low Reynolds number plasmas.   Ref. \cite{Keinigs1983} also shows that the EMF can be written  as integrals that depend  on functions of the current helicity spectra OR kinetic helicity spectra, so that net current helicity density must be finite for a finite EMF.  For our case, when shear  provides the phase correlator, there is more symmetry between kinetic and current helicity;  the two are  proportional to each other and so is the EMF.
  
That our   EMF is  proportional to either the kinetic or current helicity densities associated with fluctuations raises the question of how this result relates to conventional high $R_M$
mean field dynamos  where the $\alpha$ effect is the {\it difference} between current and kinetic helicity  densities of fluctuations ~\cite{pouquet}.  
 In these contexts, unlike ours, the kinetic helicity is often 
 driven with velocities much stronger than that of the initial mean field and the current helicity builds up to  quench the dynamo coefficient \citep{blackmanfield2002,subramanian2004nonlinear}. 
However, for large $R_M$ flows, $\alpha$ emerges as a difference in these helicities only when the nonlinear triple fluctuation terms are included and approximated with a closure \citep{pouquet,blackmanfield2002}.  
In approaches which exclude triple fluctuation terms  but apply the first order smoothing approximation (FOSA) yet still find $\alpha$ to be proportional to a difference of helicities, only the  kinetic helicity computed from the part of the fluctuations that are  strictly independent of the large scale field enters that difference.

In our present approach, the quasilinear helicity terms are constructed from  linear perturbations with $|\tilde{\bf V}|\sim |\tilde{\bf B}|$. Although we have kept the linear contribution to the Lorentz force in Eq.~\ref{eq:momentum}, the nonlinear triple fluctuation terms in the induction and velocity equations
 are absent.  The direct relationship between total kinetic and total current helicity associated with fluctuations shown in Eq.~\ref{eq:helicity2}  is therefore expected in this regime.

\textit{Magnetic helicity}:
 We started with a uniform magnetic field and no magnetic helicity, so in the absence of dissipation and boundary conditions imply that total magnetic helicity is conserved in our calculation. However,  equal and opposite large and  small scale magnetic helicities can and do arise.
 By analogy to the current helicity  that we calculated above,
 the planar averaged magnetic helicity can be written as,
$\lb\widetilde{\bf A}\cdot \widetilde{\bf B}\rb = \lb\widetilde{\bf A}\cdot \nabla \times \widetilde{\bf A}\rb = \frac{1}{2}Re\left[\frac{i (k_y^2+k_z^2)}{k_z}(\widetilde{A}_x\widetilde{ A}_y^\ast -  \widetilde{A}_y\widetilde{A}_x^\ast)\right]$.  For our system, 
$\lb\widetilde{\bf A}\cdot \widetilde{\bf B}\rb = \lb\widetilde{\bf J}\cdot \widetilde{\bf B}\rb/k^2 $. We can now write Eq (\ref{eq:emf2}) in terms of the magnetic helicity,
\begin{equation}
    \emfb \cdot \bf \bar B= \lb \widetilde {\bf V} \times \widetilde {\bf B}\rb \cdot \bf \bar B \equiv - \gamma \lb\widetilde \textbf A \cdot \widetilde \textbf B\rb.
  \label{eq:emf3}
\end{equation}
Eq. (\ref{eq:emf3}) shows that  the large-scale field generating EMF is sustained
by  helicity production at the scales of fluctuations. Were we to include 
a magnetic diffusivity term ($\eta\nabla^2 \widetilde \textbf B$) in the RHS of Eq.~\ref{eq:induction}, the factor $\gamma$ in our calculations is simply replaced by $\gamma +\eta k^2$ and the right of Eq.  (\ref{eq:emf3}) would become
 $- \gamma \lb\widetilde \textbf A \cdot \widetilde \textbf B\rb - \eta \lb\widetilde \textbf J \cdot \widetilde \textbf B\rb$. 

To check   total magnetic helicity conservation, we also calculate the magnetic helicity at large-scale (mean), $\lb{\bf \bar A}\cdot{\bf \bar B}\rb \equiv (\bar A_z\bar B_0) $. Using $\overline B_y=-\frac{\partial \overline A_z}{\partial x}$, and $\overline B_{y} = - {1\over \gamma} \frac{\partial \emf_z}{\partial x}$, \begin{equation}
    \emfb \cdot \bf \bar B= \lb \widetilde {\bf V} \times \widetilde {\bf B}\rb \cdot \bf \bar B \equiv \gamma \lb \bf \bar A \cdot \bf \bar B\rb.
  \label{eq:emf4}
\end{equation}
Thus, based on  direct calculation of EMF and magnetic helicity, Eqs ~\ref{eq:emf3}-\ref{eq:emf4} imply that the total magnetic helicity $\lb \bf  A \cdot \bf B\rb_{\mathrm{total}} = \lb \bf \bar A \cdot \bf \bar B\rb + \lb{\widetilde \textbf A \cdot \widetilde \textbf B}\rb =0$  in our system.

\textit{Broad implications:} 
There is substantial evidence for large scale field growth in local shearing boxes with and without rotation, and global simulations \citep{Brandenburg1995,Yousef2008,brandenburg2008,ebrahimi2009,gressel2010,Lesur2010,stratified2010,simon2011,Guan2011,Sorathia2012,ebrahimiprl,Suzuki2014,EB_2016,bhat2016}. Not without controversy, attempts   to explain  large scale dynamo sustenance after MHD turbulence has saturated have previously employed   (i) conventional $\alpha-\Omega$  dynamo equations \citep{Brandenburg1995,simon2011,Guan2011} with empirically extracted dynamo coefficients from the simulations (ii)  shear directly amplifying non-helical turbulence \cite{Blackman1998} or fluctuating $\alpha$-effect dynamos \cite{vishniac97,Heinemann2011,sridhar2014} or (iii)   shear-current effects \cite{shearcurrent,Kleeorin2003+,Squire2015+}.
The mechanism obtained here, shear-driven $\alpha$-effect, however, is distinct from the models listed  above.

Although our present work focuses on a regime where  perturbations are small compared to the mean field and without rotation,   we can extract  several lessons
 that apply more broadly to the  non-perturbative regime. 
  First,  our shear-plane  averaging seems   more appropriate 
for astrophysical flows than the commonly used $x-y$ averaging in shearing box simulations when comparing to standard mean field  axisymmetric accretion disk theory and spectral calculations, which  leave quantities as a function of radius ($x$ in our Cartesian coordinates) not $z$. 

Second,  even in the context of $x-y$ averaging,  our results exemplify that any approach restricted  to   spatially uniform $\alpha$ coefficients  cannot reveal a dynamo  that depends on   spatial variation of $\alpha$. This also means that no term in the EMF proportional to a spatial derivative of the mean field  is required for initial growth in our shear driven $\alpha$ effect dynamo.
A spatial variation of $\alpha$ may be important even when its average over the remaining  coordinate vanishes. The latter  has been   interpreted to suggest that $\alpha$ is  less important than e.g. a negative  diffusion tensor component  in shearing box simulations, even though both are present \cite{Shi+2016}. But a spatially varying $\alpha$ may be an essential part of 
the underlying mechanism 
which leads to ambiguity as to the dominant effect sustaining the associated large scale fields.
Self-consistent mean field theory should invoke the same averaging procedure for dynamo coefficients like $\alpha$ as for $\overline{ \bf B}$ itself.  If the latter is planar averaged then the dynamo coefficients  should  be as well.  This highlights an inconsistency in previous calculations \cite{Guan2011, 
Squire2015+,Shi+2016}  that  use volume averages for  dynamo coefficients and  planar averages for fields.

In short, 
we have uncovered  a new shear-driven $\alpha$-effect that explicitly depends on position for
shear-plane averages, and  on
 the shear for the minimalist conditions considered (Eq.~\ref{eq:alpha}). This would not be discovered if homogeneity of the dynamo coefficients were assumed a priori, and requires a correct averaging procedure whereby  mean fields and dynamo coefficients are  averaged in the same way. 
In the long-standing effort to  understand  large scale field growth and its connection to
the magnetorotational instability 
in sheared rotators, the presently derived shear-plane averaged, shear-driven $\alpha$-effect dynamo warrants further consideration.

\acknowledgements

FE acknowledges support by DOE grant DE-AC02-09CHI1466.

\end{document}